\documentclass[10pt,twocolumn,letterpaper]{article}

\usepackage{iccv}
\usepackage{times}
\usepackage{epsfig}
\usepackage{graphicx}
\usepackage{amsmath}
\usepackage{amssymb}
\usepackage{multirow}
\usepackage{caption}
\usepackage{xcolor}
\usepackage{authblk}

\usepackage[breaklinks=true,bookmarks=false]{hyperref}

\iccvfinalcopy 

\ificcvfinal\fi

\begin{document}

\title{HyperCoil-Recon: A Hypernetwork-based Adaptive Coil Configuration Task Switching Network for MRI Reconstruction\vspace{-2em}}
\author[1,2]{Sriprabha Ramanarayanan}
\author[1]{Mohammad Al Fahim}
\author[1,2]{Rahul G. S.}
\author[1]{Amrit Kumar Jethi}
\author[2]{Keerthi Ram}
\author[1,2]{Mohanasankar Sivaprakasam}
\affil[1]{Indian Institute of Technology Madras (IITM), India}
\affil[2]{Healthcare Technology Innovation Centre, IITM, India}

\maketitle
\ificcvfinal\thispagestyle{empty}\fi
\begin{abstract}
Parallel imaging, a fast MRI technique, involves dynamic adjustments based on the configuration \ie number, positioning, and sensitivity of the coils with respect to the anatomy under study. Conventional deep learning-based image reconstruction models have to be trained or fine-tuned for each configuration, posing a barrier to clinical translation, given the lack of computational resources and machine learning expertise for clinicians to train models at deployment. Joint training on diverse datasets learns a single weight set that might underfit to deviated configurations. We propose, HyperCoil-Recon, a hypernetwork-based coil configuration task-switching network for multi-coil MRI reconstruction that encodes varying configurations of the numbers of coils in a multi-tasking perspective, posing each configuration as a task. The hypernetworks infer and embed task-specific weights into the reconstruction network, 1) effectively utilizing the contextual knowledge of common and varying image features among the various fields-of-view of the coils, and 2) enabling generality to unseen configurations at test time. Experiments reveal that our approach 1) adapts on the fly to various unseen configurations up to 32 coils when trained on lower numbers (\ie 7 to 11) of randomly varying coils, and to 120 deviated unseen configurations when trained on 18 configurations in a single model, 2) matches the performance of coil configuration-specific models, and 3) outperforms configuration-invariant models with improvement margins of $\sim$ 1 dB / 0.03  and 0.3 dB / 0.02 in PSNR / SSIM for knee and brain data. Our code is available at https://github.com/sriprabhar/HyperCoil-Recon
\end{abstract}
\vspace{-0.5cm}
\section{Introduction}

\begin{figure*}[t]
    \centering
    \includegraphics[width=0.8\linewidth]{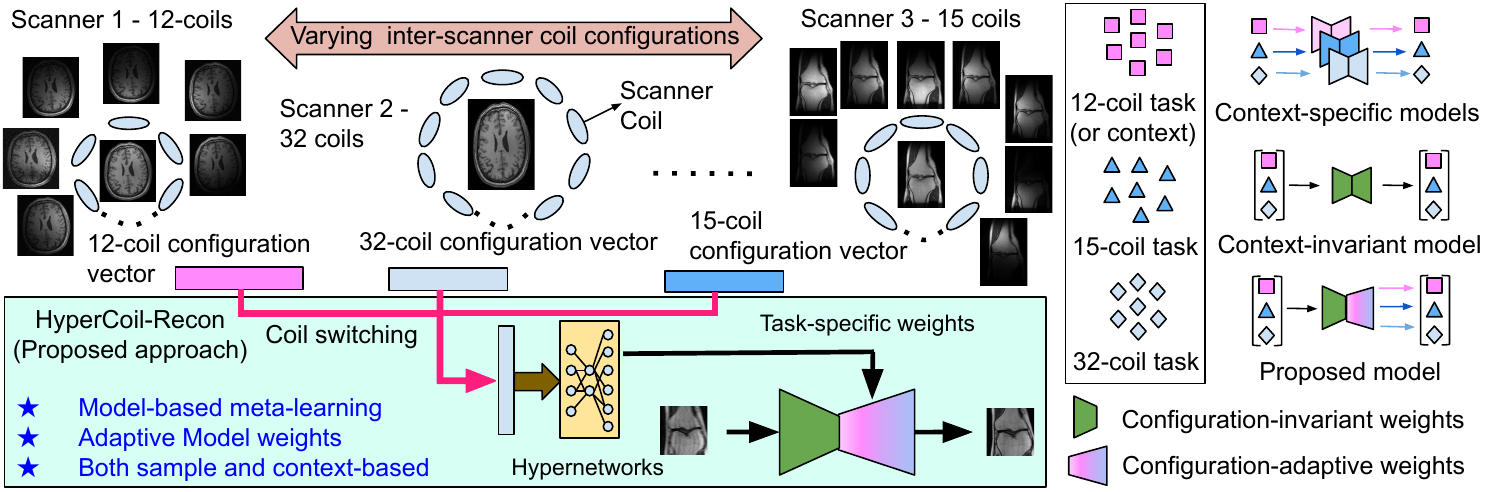}
    \caption{ The hypernetwork-based coil-configuration task switching model for adaptive MRI reconstruction. Task-specific models need training for every coil configuration, while joint training has single shared weights, HyperCoil-Recon infers task-adaptive weights for the reconstruction network, enabling generalization to several unseen contexts without retraining.
    }
    \label{fig:concept}
\end{figure*}
Parallel Imaging (PI) is a widely used technique in reducing the acquisition time of Magnetic Resonance Imaging (MRI) \cite{grappanewmethod}. Modern MRI scanners adopt PI as the default option to image the patient’s anatomy by acquiring frequency-domain (or k-space) measurements  using an instrument called a receiver coil. They employ multiple receiver coils that simultaneously obtain under-sampled k-space data of different views of the anatomy being imaged to speed up the acquisition \cite{multicoilmidl2022}. The main characteristics of PI are: 1) The captured views change for each scan and are dependent on the configuration (number and positioning) of the coils with respect to the anatomy being imaged \cite{grappanet}. 2) Different coils are typically sensitive to different but overlapping regions based on their interaction with the anatomy \cite{grappanet}. These factors indicate that PI is not only diversified by the multimodal nature of the MRI data \cite{unifiedhypergan,kmmaml}, (\eg various contrasts) but also characterized by dynamic contextual adjustments in the coil configurations at scan time.
Recently, deep learning methods have shown promising results over conventional methods like SENSE \cite{sense} and GRAPPA \cite{grappa} to reconstruct images from under-sampled multi-coil k-space owing to their capabilities to learn complex representations from the data \cite{mri_survey}. However, despite their success, standard deep neural networks (DNNs) remain restrictive under two training conditions. 

1. \textbf{Context-invariant or joint training: } Conventional DNN models, when presented with images from diverse acquisition contexts while training, employ a naive data-loading logic that combines and shuffles the images from the contexts. 
This joint training process learns a fixed set of weights that contain features common across all the contexts considered during training \cite{ledig, univusmri} (Figure \ref{fig:concept}).
However, due to the diversity and complementarity of the multi-coil MRI, different coil configurations representing different numbers of coils, exhibit heterogeneous visual characteristics. As a result, the shared features learned using a single set of weights could underfit contexts that are deviated from the training data during inference \cite{fastnflexible}. For instance, the MIDL Calgary Campinas Multi-coil MRI reconstruction challenge \cite{calgarymc} takes a clinical scenario of adding more coils in a scanner and raises concern about the generalizability of the DNN reconstruction models to the 32-coil dataset, when trained on the 12-coil dataset for a given anatomy. It is noted that 28\% of the images with unseen coil configurations at test time, assessed by expert observers were deemed to have poorer image quality when compared with the performance of the trained reference model \cite{calgarymc}. 

2. \textbf{Context-specific training: } The architectures used to solve the MRI reconstruction tasks are often similar.  Yet, the models have to be trained separately for a dataset with different acquisition settings (different coil configuration, anatomies, and contrasts) \cite{neuralizer,mac} (Figure \ref{fig:concept}). Methods like test-time training adapt the model to the target distribution by back-propagation using an unsupervised loss on complete test data distribution \cite{ontheflytta}.  These approaches might be infeasible for clinical translation, given the lack of large computational resources or machine-learning expertise for clinicians to train models at deployment time \cite{neuralizer}. 
As PI involves overlapping fields-of-view of different coils for a given scan \cite{coilmri}, different coil-switching configurations obtained from the presence or absence of various coils carry both common and varying image domain features among them. Utilizing this knowledge motivates the need for a meta-model that dynamically changes across varying coil configurations, enabling generality to new configurations in a zero-shot setting, without model updates at deployment.

Inspired by hypernetworks \cite{hypernetworks,conthyp}, a recently emerged deep learning technique in providing adaptability,  information sharing, and data-efficiency in DNNs for multi-tasking \cite{hypernet_review},  we propose a clinically motivated setting for multi-coil MRI reconstruction, called the coil-configuration task switching neural network. We pose each coil configuration as a task that encodes varying coil-switching combinations \ie  the presence and absence of one or more coils in the multi-coil MRI data. 
Our approach employs small task-conditioned hypernetworks as meta-learners \cite{meta_mihi} that infer task-specific weights and embed these latent representations of various coil configurations into an encoder-decoder-based network for MR image reconstruction (Figure \ref{fig:concept}). The encoder is task-invariant and learns shared features across coil configurations while the decoder is task-adaptive. The hypernetworks 1) infuse the coil configuration-specific embeddings in all layers of the decoder to integrate both task-specific and task-invariant features, 2) offer interesting insight into the relationship between various coil-configuration tasks, and 3) enable the model to adapt, on the fly, to various unseen coil configurations across contrasts, anatomies, and datasets in a single forward pass during inference. Our contributions are,

1. We propose HyperCoil-Recon, an adaptive coil configuration task-switching network for test-time on-the-fly adaptation to varying coils in multi-coil MR image reconstruction. The proposed method adopts a multi-tasking approach by posing each coil configuration as a task and learning from both task embeddings and under-sampled images.

2. As the first endeavor to facilitate task generalization in multi-coil MRI reconstruction, we focus on analyzing the capabilities of such a unified model when combining multiple contrasts and anatomies and presenting insights into the relationship between various coil configurations.

3. Our method uses hypernetworks as meta-learners to enrich the training process and embeds coil-configuration-specific information into the reconstruction network.

4. Our experiments reveal that our approach (i) adapts on the fly to various unseen configurations up to 32 coils when trained on lower numbers (\ie 7 to 11) of randomly varying coils, generalizes to 120 deviated unseen  configurations when trained on 18 configurations, in a single unified model, (ii) matches the performance of coil configuration-specific models, and (iii) outperforms  configuration-invariant models with improvement margins of $\sim$ 1 dB / 0.03  and 0.3 dB / 0.02 in PSNR / SSIM for knee and brain anatomies, respectively.

\section{Related Work}

\noindent\textbf{Hypernetworks:} Hypernetworks are neural networks that generate weights for another neural network, known as the target or primary task-oriented network. 
Hypernetworks have shown promising results in a variety of deep learning problems, including transfer learning \cite{oneshothyp}, continual \cite{conthyp} and meta-learning \cite{mmaml}, causal inference \cite{causalinf}, and neural architecture search \cite{oneshothyp}. In image restoration tasks, hypernetworks are used in decoupled learning \cite{gnldecouple} to solve multiple parameterized imaging operators and controllable image restoration using a single  network \cite{controllable,cfs-net}.  
The decoupled learning methods lack the inductive bias of convolution layers of the primary network to solve the imaging tasks, as the hypernetworks are the main source of weights for the primary network.  In our approach, both the primary network and the hypernetworks 
are jointly learned, enabling both coil context-invariant image features and context-specific semantic features.  
The controllable image processing networks adopt hypernetworks as a tuning module to adjust the input parameters of the imaging operator (\eg scale factor in image super-resolution). Different from this approach which uses a single parameter, in our work, the hypernetworks are conditioned with comprehensive task embeddings of the coil switching to infer task-specific weights.  The hypernetworks have non-linear leaky ReLU layers, which prevent zero activations for negative layer weights, enabling expressivity to numerous contexts.
 
\noindent\textbf{Adaptive MRI reconstruction networks:} 
Several task-specific convolutional neural networks (CNNs) have been developed for MRI reconstruction, namely unrolled networks \cite{dc_unet, dc_wcnn, dc_cnn}, attention mechanism \cite{miccan}, transformer-based networks \cite{huang2022swin}, and  variational networks \cite{yiasemis2022recurrent}. Adaptive single-coil reconstruction networks include, (i) MAC-ReconNet \cite{mac}, and Hyper-Recon \cite{wang2021regularizationagnostic}, based on decouple learning, wherein the hypernetworks are driven based on scanner information and regularization hyper-parameters, respectively, and (ii) the universal under-sampled MRI reconstruction \cite{univusmri} and side-information guided networks \cite{mrirecon_sideinfo}, based on adaptive instance normalization (AdaIN) \cite{taskswitch}. 
 MAC-ReconNet uses layer-wise linear hypernetworks with limited-sized input layers for a deeper base network. In ours, we adopt wider non-linear hypernetworks to learn task relationships and they are not heavy due to the presence of a hidden bottleneck embedding layer. The number of hypernetworks depends only on the sub-sampling levels and not on the number of layers in the primary network. 
Unlike AdaIN, the hypernetworks in our method are more expressive \cite{infinitewidthhypernetworks} due to dense multiplicative interactions \cite{MultiplicativeIA} via convolution of the dynamic weights with the CNN features. 

\section{Method}

\begin{figure*}[t]
    \centering
    \includegraphics[width=\linewidth]{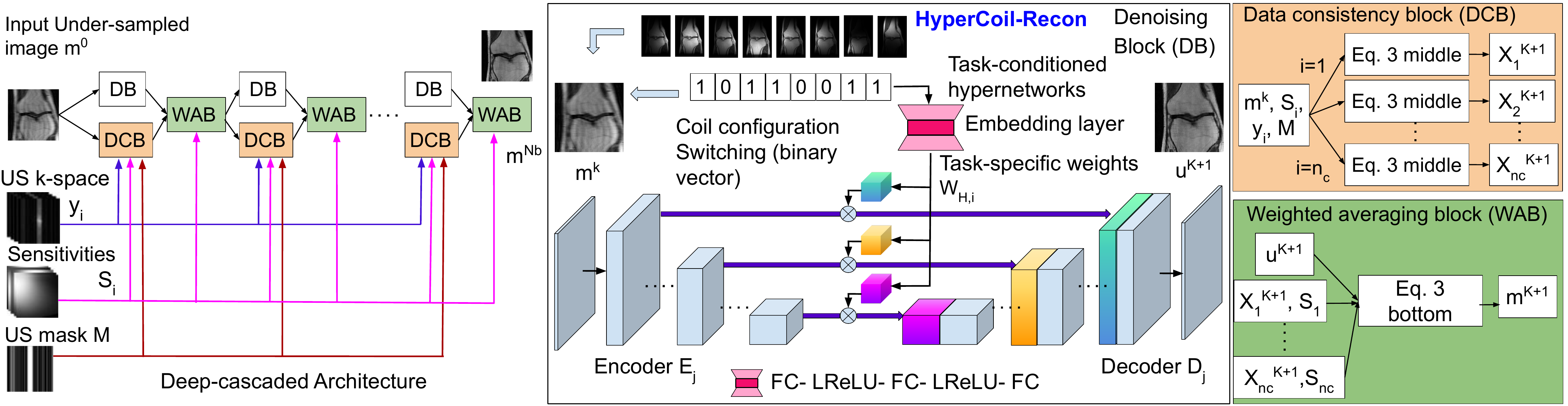}
    \caption{Deep Cascaded HyperCoil-Recon Architecture with the denoising  (the hypernetworks and the reconstruction network), data consistency, and weighted-average blocks. FC - Fully connected layer, LReLU - Leaky ReLU non-linearity.
    }
    \label{fig:arch}
\end{figure*}

\textbf{Problem Formulation:} According to compressed sensing for MRI, the problem of recovering the desired complex-valued MR image, $x$ $\in$ $\mathbb{C}^\emph{N}$, from the under-sampled (US) k-space measurements, $y$ $\in$ $\mathbb{C}^\emph{M'}$ is ill-posed as $M'<<N$ \cite{dcrsn} and the optimization formulation is,

\begin{equation}
 \label{Optimization formulation}
 \underset{x} {\operatorname{min}} \quad  \sum_{j=1}^{n_c} ||MFS_{j}x - y_{j}||_{2}^2 + \mathcal{R}(x)
\end{equation}

Here, $\mathcal{R}(x)$ is the sparse regularization term, $M$, the 2-D under-sampling mask, $F$, the Fourier transform matrix, and $S_j$, $j = 1, 2, .., n_{c}$, the sensitivity maps of the $n_{c}$ receiver coils. 
The proposed deep learning-based MRI reconstruction involves training a deep learning (DL) model on the average loss of all observed data samples from $N_{\gamma}$ datasets. Each dataset, corresponding to the coil configuration task $\gamma_{i}$, consists of US or zero-filled (ZF) image input, and the fully-sampled (FS) image target pairs denoted as $D_{i} = (x_{US,i}, x_{FS,i})$, where $i = 1, 2, .., N_{\gamma}$ is the task index.
This supervised task-aware joint training of the neural network, $f(x_{US,i}, \overrightarrow\gamma_{i};\theta,\phi_{H})$ consisting of the image reconstruction CNN, $f_{CNN}(x_{US};\theta)$ with parameters $\theta$ and the hypernetwork, $f_{H}(\overrightarrow\gamma_{i};\phi_{H})$ with parameters $\phi$, across multi-coil configuration tasks, $\gamma_{i}$ is given by,

\begin{equation}
 \label{metaMR}
 \begin{split}
 \theta^{*}, \phi^{*}_{H} = \underset{\theta, \phi_{H}} {\operatorname{argmin}}  \displaystyle \mathop{\mathbb{E}}_{(x_{US,i},x_{FS,i})\in \bigcup\limits_{i}^{} \mathcal{D}_{i}}[||x_{FS,i} - \\ f(x_{US,i},\overrightarrow\gamma_{i};\theta, \phi_{H}) ||_{2}^2]
 \end{split}
\end{equation}

Here, $\theta^{*}$ and $\phi^{*}_{H}$ are the optimized weights of the networks. The configuration vector input $\overrightarrow\gamma_{i}$ to the hypernetwork is a binary vector that enumerates the presence  (binary 1) and absence (binary 0) of a coil (\ie the corresponding sensitivity map and the k-space).
In order to optimize Eq. \ref{Optimization formulation} efficiently, the variable splitting method in VS-Net \cite{vs_net} introduces auxiliary splitting variables $u \in C^{N}$ and $\{x_{j} \in C^{N}\}^{n_{c}}_{j=1}$ and derives the final solution as follows:

\begin{equation} 
\label{eq:problem_formulation_vs}
\footnotesize
\begin{split}
u^{k+1} &= denoiser(m^{k})  \\ 
x_{j}^{k+1} &= F^{-1}((\lambda M^{T}M + \alpha I)^{-1}(\alpha FS_{j}m^{k} + \lambda M^{T}y_{j})) \\ 
m^{k+1} &= (\beta I + \alpha \sum_{j=1}^{n_{c}}S_{j}^{H}S_{j})^{-1}(\beta u^{k+1} + \alpha \sum_{j=1}^{n_{c}}S_{j}^{H}x_{j}^{k+1})
\end{split}
\end{equation}

The top equation converts the original problem (Eq. \ref{Optimization formulation}) to a denoising problem as in Eq. \ref{metaMR}. The middle equation provides the data consistency to k-space for each coil. The bottom equation computes a weighted average of the results obtained from the first two equations. 
The proposed model (Fig. \ref{fig:arch}) for multi-coil MRI reconstruction follows the iterative setup formulated in Eq. \ref{eq:problem_formulation_vs} with $N_{b}$ cascades of the three blocks: HyperCoil-Recon as denoiser block, data consistency block (DCB) and weighted average block (WAB). The DL model takes in 
 the sensitivity-weighted US image ($m^{0} =  \sum_{j=1}^{n_{c}}S_{j}^{H}F^{-1}M^{T}y_{j}$) and it's corresponding binary configuration vector 
 as inputs. DCB uses the binary sampling mask ($M$), the under-sampled k-space data ($\{M^{T}y_{j}\}^{n_{c}}_{j=1}$) with the corresponding sensitivity maps for which the coil configuration switching vector has binary 1 value to provide k-space data consistency. 
WAB uses coil sensitivity maps to perform weighted averaging of the DCB outputs and combine it with the DL model output. 

\noindent \textbf{Architecture Details:}
The primary reconstruction CNN is the U-Net \cite{unet}. 
There are $N_{sub}$ hypernetworks that take the coil configuration task vector as input and infer task-specific weights $W_{H, i}$ corresponding to the $N_{sub}$ sub-sampling levels of the decoder in the primary CNN. The encoder features $E_{1}, E_{2}, ..., E_{N_{sub}}$ of the CNN are convolved (shown as $\otimes$) with the $N_{sub}$ task-specific weight vectors to obtain task-specific features, that are concatenated with the decoder features $D_{1}, D_{2}, ..., D_{N_{sub}}$ at each level and the reconstructed image $x_{CNN, i}$ for the task, $\gamma_{i}$, is obtained (Fig. \ref{fig:arch}).

\begin{equation}
\begin{split}
W_{H, i} = (W_{H,1,i}, ..., W_{H,N_{sub},i}) =  f_{H}(\overrightarrow\gamma_{i};\phi_{H}) \\
F_{Dec, j, i} = (E_{j} \otimes W_{H,j,i})\ ||\ D_{j}, 
\forall\  j = 1, ...,  N_{sub} \\
x_{CNN, i} = Dec(F_{Dec, 1, i}  , ... , F_{Dec, N_{sub}, i})
\end{split}
\label{csweights}
\end{equation}



\section{Experiments}

\subsection{Datasets and Implementation Details}

\noindent\textbf{Datasets:} 1) The \textbf{Knee multi-coil dataset} \cite{variational} consists of coronal proton-density without (PD) and with fat suppression (PDFS) images, acquired from 20 patients. 
Each patient data has 20 slices of size 640 $\times$ 368 with 15 channels (coils) and sensitivity maps. 
We split the data into 10 slices with 200 slices each, for training and validation with Cartesian under-sampling. 2) \textbf{Calgary Campinas Challenge multi-coil dataset} \cite{calgarymc}: This large-scale publicly accessible dataset provides k-space data from 167 3D, T1-weighted, gradient-recalled echo, 1 $mm^3$ isotropic sagittal brain scans collected on a clinical 3-T MRI scanner (Discovery MR750; GE Healthcare). There are two datasets, one with 12-channel (117 scans) and the other with 32-channel (50 scans) receiver coils, each with 170 to 180 contiguous slices of size $256 \times 218$. The dataset consists of Poisson under-sampling masks for 5x and 10x acceleration.

\noindent\textbf{Implementation Details: }
We represent each coil configuration task vector as a binary vector that encodes the coil switching. For example, in a 15-coil dataset, a 9-coil  task switching vector, 111001010101110 denotes that the sensitivities and k-space data of coils 4, 5, 7, 9, 11, and 15 are absent while the rest are present.  For a given number of coils (task),  7, 9, 11, and 12, we augment the task with several randomly varying combinations (sub-tasks) of binary-valued vectors, embed them in a 32-bit vector (maximum embedding vector size) initialized with 1's and feed to the hypernetworks to support multiple configurations in one training. We have implemented the models using Pytorch v1.12, trained for 100 epochs with 5 cascades on a 24 GB Nvidia RTX-3090 and L1 as the loss function between the predicted and the ground truth (GT) fully sampled images. Our evaluation metrics are PSNR and SSIM measures.

\subsection{Results and discussion}
Our experiments include, 1) Generalization to unseen coil configurations when trained on few configurations, 2) Task relationship, 3) Performance Comparison with other multi-coil MRI reconstruction architectures on large-scale clinical datasets, 4) Comparison with other adaptive MRI reconstruction methods for multi-modal acquisition contexts, and 5) An ablative study. 

\begin{table*}[t!]
\centering
\scriptsize
\caption{Quantitative comparison of the generality of the proposed model for unseen configurations with CCTIM and CCTSM when combining 7, 10, \& 12 coils, 4, 5, \& 8x acceleration, and contrasts - PD \& PDFS. ZF - Zero-filled (\ie US) image.
For \eg, the task PD7 - proton density MRI with 7x acceleration. Green and blue - the first and second-best metrics, respectively.}
\label{tab:table1}
\begin{tabular}{|c|c|c|c|c|c|c|c|c|c|c|c|}
\hline
                        &                                                                              & ZF                                                    & CCTIM                                                 & \begin{tabular}[c]{@{}c@{}}HyperCoil-\\ Recon\end{tabular} & CCTSM                                                 &                         &                                                                              & ZF                                                    & CCTIM                                                 & \begin{tabular}[c]{@{}c@{}}HyperCoil-\\ Recon\end{tabular} & CCTSM                                                 \\ \cline{3-6} \cline{9-12} 
\multirow{-2}{*}{Coils} & \multirow{-2}{*}{\begin{tabular}[c]{@{}c@{}}Scanner \\ context\end{tabular}} & \begin{tabular}[c]{@{}c@{}}PSNR/ \\ SSIM\end{tabular} & \begin{tabular}[c]{@{}c@{}}PSNR/ \\ SSIM\end{tabular} & \begin{tabular}[c]{@{}c@{}}PSNR/ \\ SSIM\end{tabular}      & \begin{tabular}[c]{@{}c@{}}PSNR/ \\ SSIM\end{tabular} & \multirow{-2}{*}{Coils} & \multirow{-2}{*}{\begin{tabular}[c]{@{}c@{}}Scanner \\ context\end{tabular}} & \begin{tabular}[c]{@{}c@{}}PSNR/ \\ SSIM\end{tabular} & \begin{tabular}[c]{@{}c@{}}PSNR/ \\ SSIM\end{tabular} & \begin{tabular}[c]{@{}c@{}}PSNR/ \\ SSIM\end{tabular}      & \begin{tabular}[c]{@{}c@{}}PSNR/ \\ SSIM\end{tabular} \\ \hline
                        & PD7                                                                          & 18.47/ .717                                           & 28.64/ .780                                           & {\color[HTML]{009901} 29.38/ .821}                         & {\color[HTML]{3531FF} 27.93/ .768}                    &                         & PD7                                                                          & 25.44/ .822                                           & 30.80/ .799                                           & {\color[HTML]{009901} 31.53/ .862}                         & {\color[HTML]{3531FF} 30.88/ .845}                    \\ \cline{2-6} \cline{8-12} 
                        & PD9                                                                          & 18.46/ .714                                           & 28.28/ .770                                           & {\color[HTML]{009901} 29.03/ .814}                         & {\color[HTML]{3531FF} 27.58/ .756}                    &                         & PD9                                                                          & 25.32/ .816                                           & 30.35/ .791                                           & {\color[HTML]{009901} 31.00/  .853}                        & {\color[HTML]{009901} 30.44/ .837}                    \\ \cline{2-6} \cline{8-12} 
                        & PDFS7                                                                        & 21.91/ .602                                           & 29.85/ .733                                           & {\color[HTML]{009901} 30.30/ .746}                         & {\color[HTML]{3531FF} 29.76/ .732}                    &                         & PDFS7                                                                        & 28.30 / .760                                          & 31.90/ .766                                           & {\color[HTML]{009901} 32.39/ .789}                         & {\color[HTML]{3531FF} 32.07/ .782}                    \\ \cline{2-6} \cline{8-12} 
\multirow{-4}{*}{7}     & PDFS9                                                                        & 21.89/ .598                                           & 29.56/ .726                                           & {\color[HTML]{009901} 30.00/ .740}                         & {\color[HTML]{3531FF} 29.49/ .725}                    & \multirow{-4}{*}{13}    & PDFS9                                                                        & 28.20/ .753                                           & 31.52/ .757                                           & {\color[HTML]{009901} 32.04/ .781}                         & {\color[HTML]{3531FF} 31.73/ .774}                    \\ \hline
                        & PD7                                                                          & 20.17/ .761                                           & 29.61/ .798                                           & {\color[HTML]{009901} 30.20/ .835}                         & {\color[HTML]{3531FF} 30.19/ .828}                    &                         & PD7                                                                          & 27.01/ .823                                           & 30.87/ .798                                           & {\color[HTML]{3531FF} 31.52}/ {\color[HTML]{009901}.861}                         & {\color[HTML]{009901} 31.79}/ {\color[HTML]{3531FF}.847}                    \\ \cline{2-6} \cline{8-12} 
                        & PD9                                                                          & 20.14/ .757                                           & 29.21/ .789                                           & {\color[HTML]{3531FF} 29.78}/ {\color[HTML]{009901}.828}                         & {\color[HTML]{009901} 29.79/ .818}                    &                         & PD9                                                                          & 26.82/ .823                                           & 30.41/ .789                                           & {\color[HTML]{3531FF} 30.99}/ {\color[HTML]{009901}.852}                         & {\color[HTML]{009901} 31.25}/ {\color[HTML]{3531FF}.838}                    \\ \cline{2-6} \cline{8-12} 
                        & PDFS7                                                                        & 23.57/ .665                                           & 30.90/ .752                                           & {\color[HTML]{3531FF} 31.23/ .766}                         & {\color[HTML]{009901} 31.27/ .768}                    &                         & PDFS7                                                                        & 29.75/ .775                                           & 31.93/ .766                                           & {\color[HTML]{3531FF} 32.46}/ {\color[HTML]{009901}.790}                         & {\color[HTML]{009901} 32.58}/ {\color[HTML]{3531FF}.784}                    \\ \cline{2-6} \cline{8-12} 
\multirow{-4}{*}{9}     & PDFS9                                                                        & 23.54/ .660                                           & 30.59/ .746                                           & {\color[HTML]{3531FF} 30.86/ .759}                         & {\color[HTML]{009901} 31.00/ .760}                    & \multirow{-4}{*}{14}    & PDFS9                                                                        & 29.60/ .768                                           & 31.55/ .757                                           & {\color[HTML]{3531FF} 32.01}/ {\color[HTML]{009901}.784}                         & {\color[HTML]{009901} 32.17}/ {\color[HTML]{3531FF}.776}                    \\ \hline
                        & PD7                                                                          & 22.44/ .797                                           & 30.41/ .800                                           & {\color[HTML]{009901} 30.83/ .843}                         & {\color[HTML]{3531FF} 30.37/ .818}                    &                         & PD7                                                                          & 28.88/ .836                                           & 30.87/ .797                                           & {\color[HTML]{3531FF} 31.35}/ {\color[HTML]{3531FF}.861}                         & {\color[HTML]{009901} 32.29}/ {\color[HTML]{009901}.885}                    \\ \cline{2-6} \cline{8-12} 
                        & PD9                                                                          & 22.38/ .792                                           & 29.98/ .792                                           & {\color[HTML]{009901} 30.32/ .833}                         & {\color[HTML]{3531FF} 29.95/ .810}                    &                         & PD9                                                                          & 28.61/ .829                                           & 30.42/ .788                                           & {\color[HTML]{3531FF} 30.81}/ {\color[HTML]{3531FF}.852}                         & {\color[HTML]{009901} 31.68}/ {\color[HTML]{009901}.876}                    \\ \cline{2-6} \cline{8-12} 
                        & PDFS7                                                                        & 25.75/ .719                                           & 31.59/ .762                                           & {\color[HTML]{009901} 31.90/ .778}                         & {\color[HTML]{3531FF} 31.56/ .761}                    &                         & PDFS7                                                                        & 31.20/ .786                                           & 31.90/ 0.764                                          & {\color[HTML]{3531FF} 32.43/ .793}                         & {\color[HTML]{009901} 32.91/ .803}                    \\ \cline{2-6} \cline{8-12} 
\multirow{-4}{*}{11}    & PDFS9                                                                        & 25.69/ .713                                           & 31.25/ .755                                           & {\color[HTML]{009901} 31.50/ .771}                         & {\color[HTML]{3531FF} 31.25/ .753}                    & \multirow{-4}{*}{15}    & PDFS9                                                                        & 31.00/ .779                                           & 31.50/ 0.756                                          & {\color[HTML]{3531FF} 32.04/  .785}                        & {\color[HTML]{009901} 32.46/ .794}                    \\ \hline
\end{tabular}
\end{table*}

\begin{figure*}[t!]
    \centering
    \includegraphics[width=\linewidth]{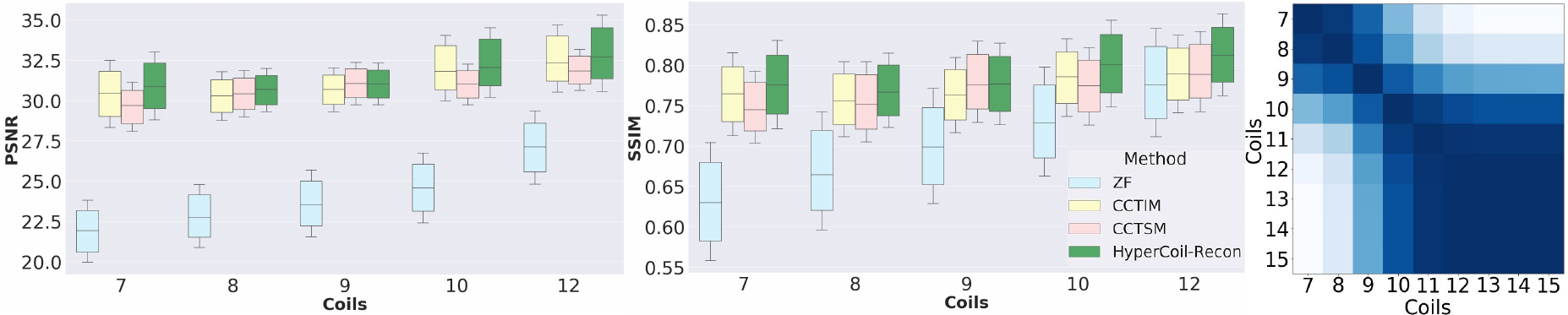}
    
    \caption{ Left \& Middle: Plots showing more unseen coil configurations when combining 7, 10, and 12 as the number of coils, acceleration factors 4x, 5x, and 8x, and contrasts - PD and PDFS. Right: Matrix plot showing the inter-task relationship. Tasks with neighboring coil configurations exhibit more similarity, while far-apart configurations exhibit lesser similarity.
    }
    \label{fig:table1plot}
\end{figure*}

\begin{figure*}[t!]
    \centering
    \includegraphics[width=0.8\linewidth]{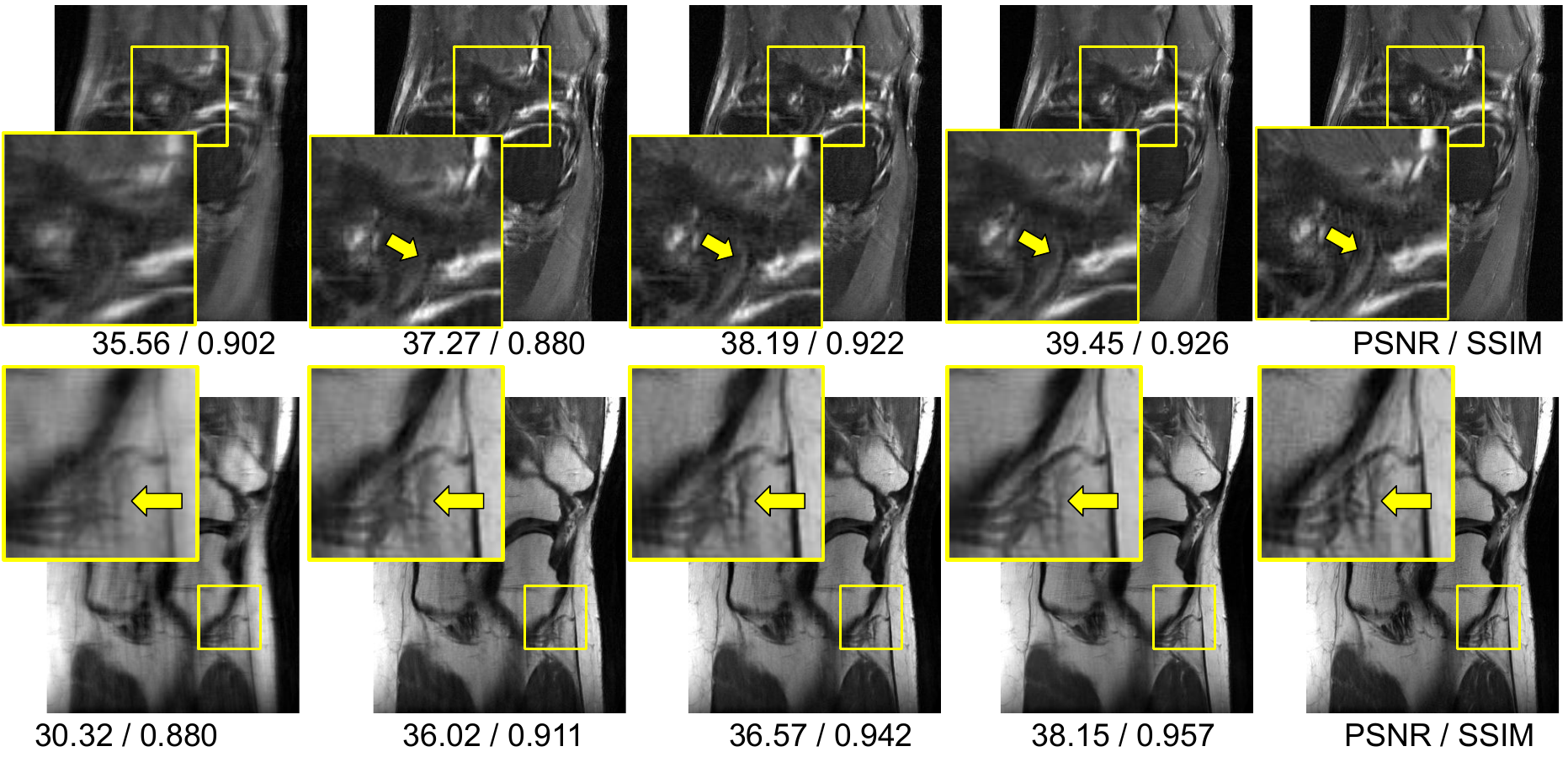}
    
    \caption{From the left: Qualitative comparison of the images of the ZF, CCTIM, HyperCoil-Recon, and CCTSM for an unseen case with 15 coils and 4x acceleration when trained on 7, 10, and 12 coils, 4x, 5x, and 8x acceleration and contrasts - PD (bottom) and PDFS (top). The quality of the proposed model is better than CCTIM and matches that of CCTSM.
    }
    \label{fig:table1}
\end{figure*}

\begin{table*}[t!]
\centering
\scriptsize
\caption{Quantitative comparison of HyperCoil-Recon with other multi-coil MRI reconstruction methods large-scale clinical data.  The column pairs are the evaluation results of the 7-9-11 model on the 12-coil unseen task (same dataset), the 7-9-11 model on the 32-coil unseen task (unseen dataset), and the 15-28-31 model on the 32-coil task (reference for column 2).}
\label{tab:table2}
\begin{tabular}{|c|cc|cc|cc|}
\hline
\multirow{3}{*}{Method} & \multicolumn{2}{c|}{\begin{tabular}[c]{@{}c@{}}12-coil testing on 7-9-11 model \\ (seen dataset \& unseen coils)\end{tabular}} & \multicolumn{2}{c|}{\begin{tabular}[c]{@{}c@{}}32-coil testing on 7-9-11 model \\ (unseen dataset and unseen coils)\end{tabular}} & \multicolumn{2}{c|}{\begin{tabular}[c]{@{}c@{}}32-coil on 15-28-31 model \\ (seen dataset \& unseen coils)\end{tabular}} \\ \cline{2-7} 
                        & \multicolumn{1}{c|}{5x}                                                       & 10x                                                                        & \multicolumn{1}{c|}{5x}                                                      & 10x                                                                       & \multicolumn{1}{c|}{5x}                                                       & 10x                                                                       \\ \cline{2-7} 
                        & \multicolumn{1}{l|}{PSNR/ SSIM}                                               & \multicolumn{1}{l|}{PSNR/ SSIM}                                            & \multicolumn{1}{l|}{PSNR/ SSIM}                                              & \multicolumn{1}{l|}{PSNR/ SSIM}                                           & \multicolumn{1}{l|}{PSNR/ SSIM}                                               & \multicolumn{1}{l|}{PSNR/ SSIM}                                           \\ \hline
ZF                      & \multicolumn{1}{c|}{25.20 / 0.726}                                            & 23.70 / 0.632                                                              & \multicolumn{1}{c|}{26.40 / 0.780}                                           & 25.50 / 0.745                                                             & \multicolumn{1}{c|}{26.40 / 0.780}                                            & 25.50 / 0.745                                                             \\ \hline
VS-Net \cite{vs_net}                  & \multicolumn{1}{c|}{27.61 / 0.862}                                            & 26.08 / 0.802                                                              & \multicolumn{1}{c|}{19.03 / 0.707}                                           & 18.61 / 0.671                                                             & \multicolumn{1}{c|}{31.04 / 0.921}                                            & 29.39 / 0.869                                                             \\ \hline
VarNet \cite{variational}                 & \multicolumn{1}{c|}{26.06 / 0.839}                                            & 24.57 / 0.789                                                              & \multicolumn{1}{c|}{19.19 / 0.673}                                           & 17.85 / 0.615                                                             & \multicolumn{1}{c|}{28.99 / 0.885}                                            & 27.49 / 0.828                                                             \\ \hline
RecurrentVarNet \cite{yiasemis2022recurrent}         & \multicolumn{1}{c|}{29.17 / 0.908}                                            & 27.46 / 0.855                                                              & \multicolumn{1}{c|}{18.81 / 0.707}                                           & 18.50 / 0.675                                                             & \multicolumn{1}{c|}{31.83 / 0.925}                                            & 30.82 / 0.893                                                             \\ \hline
KIKI-Net \cite{kikinet}                & \multicolumn{1}{c|}{31.13 / 0.911}                                            & 29.24 / 0.865                                                              & \multicolumn{1}{c|}{23.25 / 0.853}                                           & 22.50 / 0.815                                                             & \multicolumn{1}{c|}{32.44 / 0.943}                                            & 30.53 / 0.896                                                             \\ \hline
DC-Unet \cite{dc_unet}                & \multicolumn{1}{c|}{31.42 / 0.914}                                            & 29.14 / 0.865                                                              & \multicolumn{1}{c|}{23.13 / 0.859}                                           & 22.90 / 0.824                                                             & \multicolumn{1}{c|}{33.92 / 0.948}                                            & 31.72 / 0.908                                                             \\ \hline
SWin \cite{huang2022swin}                   & \multicolumn{1}{c|}{29.26 / 0.895}                                            & 27.33 / 0.834                                                              & \multicolumn{1}{c|}{19.68 / 0.756}                                           & 19.43 / 0.721                                                             & \multicolumn{1}{c|}{32.48 / 0.939}                                            & 30.39 / 0.889                                                             \\ \hline
EnchantedNet \cite{calgarymc}           & \multicolumn{1}{c|}{31.29 / 0.912}                                            & 29.00 / 0.861                                                              & \multicolumn{1}{c|}{26.63 / 0.887}                                           & 26.47 / 0.846                                                             & \multicolumn{1}{c|}{33.29 / 0.937}                                            & 30.75 / 0.889                                                             \\ \hline
HyperCoil-Recon         & \multicolumn{1}{c|}{33.54 / 0.928}                                            & 30.39 / 0.878                                                              & \multicolumn{1}{c|}{31.15 / 0.924}                                           & 29.55 / 0.882                                                             & \multicolumn{1}{c|}{36.13 / 0.955}                                            & 32.52 / 0.910                                                             \\ \hline
\end{tabular}
\end{table*}

\subsubsection{Generalization to Unseen Coil Configurations}
We train the models on a few (18) tasks and assess the generality of the models to several (120) deviated unseen tasks that share the same label space. 
From the 15-coil knee dataset, we create the 18 training tasks or configurations using combinations of  7, 10, and 12 coils, PD and PDFS contrasts, and 4x, 5x, and 8x acceleration factors. We compare our model with the jointly trained model using only the images (coil-configuration task-invariant model or CCTIM) and models trained for a specific number of coils (coil-configuration task-specific model or CCTSM). Table \ref{tab:table1} shows the performance of the models for 24 unseen coil configurations with unseen acceleration factors (7x and 9x). From the table, our observations are as follows. (i) HyperCoil-Recon consistently performs better than the CCTIM for all the coil configurations. (ii) HyperCoil-Recon outperforms CCTSM for almost all the configurations in SSIM and performs competitively in PSNR. 
(iii) In some cases, lower (\eg 13-coil) coil tasks of HyperCoil-Recon are better than higher (\eg 14-coil) coil tasks of CCTSM. 
Joint training in CCTIM lacks context awareness as the model learns a single set of shared weights from the image samples only.
Although CCTSMs operate as expert models for higher coil configurations, the knowledge gained using only the primary network is still inadequate at lower coil configurations wherein the receiving coils might not capture the anatomy of interest adequately. Furthermore, at higher acceleration factors, the CCTIM becomes comparable to ZF performance. HyperCoil-Recon benefits from 1) the low-level image features of the primary CNN, and 2) additional knowledge of context embeddings via the task-switching hypernetworks, which help to build the relationship between various contexts and enables the model to interpolate and extrapolate to unseen contexts between known contexts.  Figure \ref{fig:table1plot} shows more unseen coil configurations wherein our model generalizes better than CCTIM and CCTSM.  The visual results for HyperCoil-Recon (Figure \ref{fig:table1}) show better recovery of details over CCTIM and CCTSM.
\begin{figure*}[t!]
    \centering
    \includegraphics[width=0.9\linewidth]{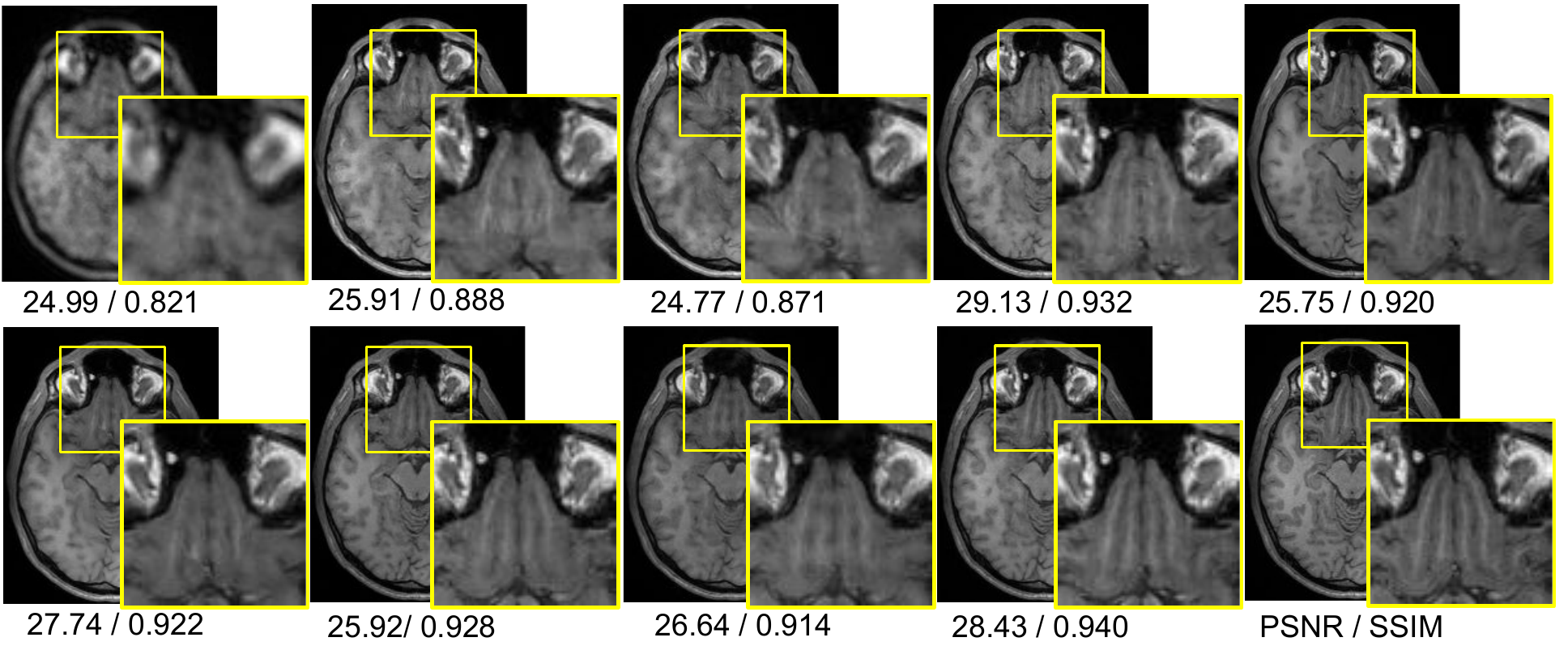}
    
    \caption{ Qualitative comparison of the HyperCoil-Recon with other multi-coil MRI reconstruction architectures for the unseen 12-coil task using the 7-9-11 model of the same dataset. Top: ZF image, VS-Net \cite{vs_net}, VarNet \cite{variational}, Recurrent VarNet \cite{yiasemis2022recurrent}, KIKI-Net \cite{kikinet}. Bottom: SWin \cite{huang2022swin}, DC-UNet \cite{dc_unet}, EnchantedNet \cite{calgarymc}, HyperCoil-Recon, and GT image.
    }
    \label{fig:table2_12coil}
\end{figure*}

\subsubsection{Understanding Task Relationships}
We interpret the role of the hypernetworks when jointly training with the primary reconstruction network with a multi-tasking objective to study if the embeddings learned by the hypernetworks carry information useful for task relationships. To analyze this perspective, we compute the similarity measure, $SIM$ between two task embedding layer output vectors $\tau_i$ and $\tau_j$ (second embedding hidden layer after LReLU in the hypernetwork shown in Fig. \ref{fig:arch}) based on cosine similarity as $SIM(\tau_{i},\tau_{j}) = 1 - \frac{\tau^T_{i}\tau_{j}}{||\tau_{i}||\ ||\tau_{j}||}$.

Figure \ref{fig:table1plot} (right) shows the matrix plot of the similarity between tasks. The darker the blue color region, the better  the similarity between the tasks (\ie $SIM$ less than 0.4) corresponding to each row-column pair. Each task denotes the number of coils used for reconstruction. We make two interesting observations. 1. The tasks that are closer in the number of coils exhibit higher similarity (\eg, Task 12 (seen) and 15 (unseen)) and those farther exhibit lesser similarity (Task 7 with Tasks 11 to 15). 2. Task similarity is denser in the region with a higher number of coils.  
We believe that when coils share common features, the tasks have common patterns in the conditioning, and the task embeddings might be localized closer to each other in the latent embedding space. This indicates that the hypernetworks share the knowledge across the tasks, facilitating adaptive coil configuration in a single unified model.  

\begin{figure*}[t!]
    \centering
    \includegraphics[width=0.95\linewidth]{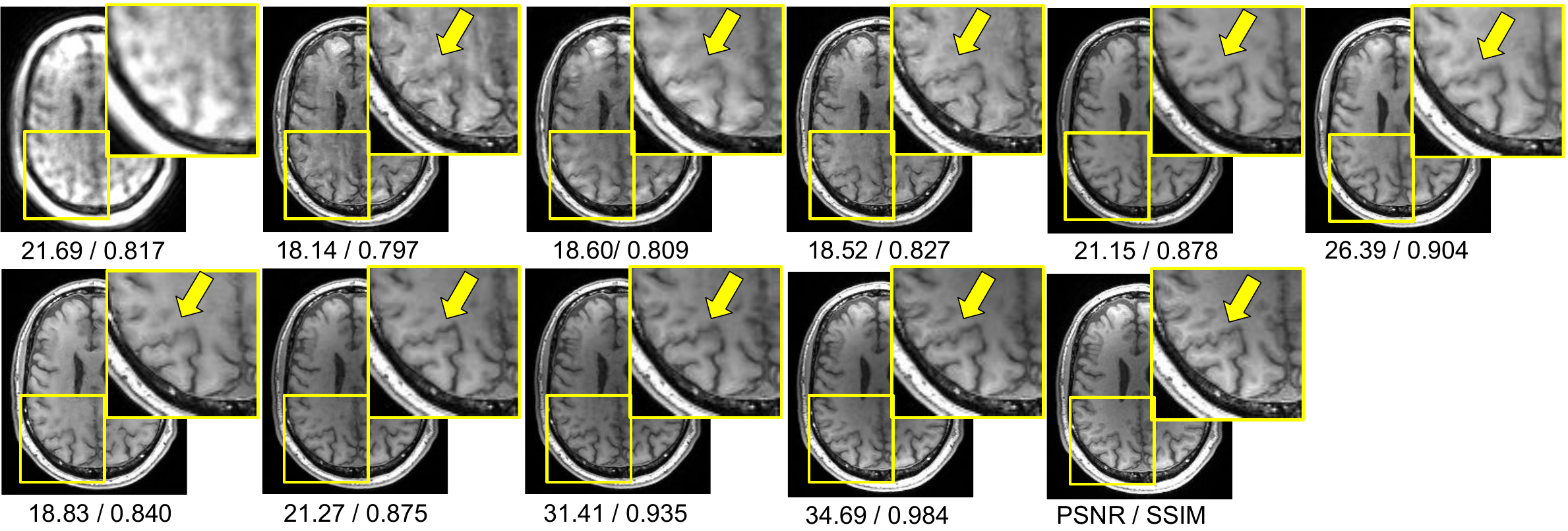}
    
    \caption{ Qualitative comparison of the HyperCoil-Recon with other methods for the unseen 32-coil Calgary dataset with unseen 32-coil configuration on the 7-9-11 model. Top: ZF image, VS-Net, VarNet, Recurrent VarNet, KIKI-Net. Bottom: SWin, DC-UNet, EnchantedNet, HyperCoil-Recon, HyperCoil-Recon (15-28-31 reference model), and GT image.
    }
    \label{fig:table2_32coil}
\end{figure*}
\subsubsection{Feature Reusability}
We evaluate the models on the Calgary-Campinas large-scale clinical  datasets of the brain anatomy. The goal of this study is to provide an objective benchmark of various multi-coil MRI reconstruction methods with respect to the generalizability of the models across the 32-coil and the 12-coil datasets.
For each reconstruction method shown in Table \ref{tab:table2}, we train two models - 1) Training with combinations of 7, 9, and 11 coils using the 12-coil Calgary dataset. 2) Training with combinations of 15, 28, and 31 coils using the 32-coil Calgary dataset. We refer to these models as the 7-9-11 and 15-28-31 models, respectively. 
Table \ref{tab:table2} shows three evaluation cases, 1) the 7-9-11 model on 12-coil unseen configuration within the same dataset,  2) the 7-9-11 model on the 32-coil unseen task on the 32-coil unseen dataset, and 3) the 15-28-31 model on 32-coil configuration within the same dataset. We take the 15-28-31 model of each method  as the reference model to assess the performance of the corresponding 7-9-11 model in Case 2.
From the table, our observations are as follows. 1) The proposed approach performs better than other reconstruction methods for all three cases. 2) The HyperCoil-Recon 7-9-11 model is comparable with the corresponding 15-28-31 model on the 32-coil task, while other models perform poorly. These observations reveal that the task-conditioned hypernetworks in the proposed model exhibit meta-learning \cite{maml} capabilities wherein the hypernetworks contain reusable features \cite{featurereuse, kmmaml} for new tasks deviated from the training tasks. Figure \ref{fig:table2_12coil} shows that the reconstructed images of our model exhibit better visual quality compared to other methods for a case with the highlighted region pointing to faint structures containing subtle image gradients. Figure \ref{fig:table2_32coil} also shows that our method gives better quality for 32 coils over other methods and is on par with the 32-coil reference  model result.
\begin{table*}[t!]
\centering
\scriptsize
\caption{Quantitative comparison of the HyperCoil-Recon with other adaptive MRI reconstruction methods - MAC-ReconNet \cite{mac} and AdaIN \cite{univusmri} under multi-modal scenario when combining (during training) different anatomies (12-coil T1 brain and 15-coil PD knee), with different contrast and different coil configurations (7, 10 and 12 coils respectively).}
\label{tab:mulanatomy}
\begin{tabular}{|c|ccc|c|ccc|}
\hline
\multirow{3}{*}{Coils} & \multicolumn{3}{c|}{Brain 12 coils}                                                       & \multirow{3}{*}{Coils} & \multicolumn{3}{c|}{Knee 15 coils}                                                        \\ \cline{2-4} \cline{6-8} 
                       & \multicolumn{1}{c|}{MAC-ReconNet \cite{mac}}  & \multicolumn{1}{c|}{AdaIN \cite{univusmri}}         & HyperCoil-Recon &                        & \multicolumn{1}{c|}{MAC-ReconNet \cite{mac}}  & \multicolumn{1}{c|}{AdaIN \cite{univusmri}}         & HyperCoil-Recon \\ \cline{2-4} \cline{6-8} 
                       & \multicolumn{1}{c|}{PSNR / SSIM}   & \multicolumn{1}{c|}{PSNR / SSIM}   & PSNR / SSIM     &                        & \multicolumn{1}{c|}{PSNR / SSIM}   & \multicolumn{1}{c|}{PSNR / SSIM}   & PSNR / SSIM     \\ \hline
7                      & \multicolumn{1}{c|}{24.95 / 0.815} & \multicolumn{1}{c|}{23.49 / 0.793} & 26.67 / 0.844   & 10                     & \multicolumn{1}{c|}{30.53 / 0.857} & \multicolumn{1}{c|}{29.44 / 0.835} & 31.98 / 0.859   \\ \hline
8                      & \multicolumn{1}{c|}{25.28 / 0.826} & \multicolumn{1}{c|}{23.88 / 0.805} & 27.04 / 0.853   & 11                     & \multicolumn{1}{c|}{31.17 / 0.860} & \multicolumn{1}{c|}{30.19 / 0.845} & 32.36 / 0.864   \\ \hline
9                      & \multicolumn{1}{c|}{25.70 / 0.838} & \multicolumn{1}{c|}{24.23 / 0.817} & 27.45 / 0.862   & 12                     & \multicolumn{1}{c|}{31.58 / 0.862} & \multicolumn{1}{c|}{30.92 / 0.853} & 32.68 / 0.868   \\ \hline
10                     & \multicolumn{1}{c|}{26.37 / 0.852} & \multicolumn{1}{c|}{24.64 / 0.827} & 27.80 / 0.869   & 13                     & \multicolumn{1}{c|}{31.89 / 0.860} & \multicolumn{1}{c|}{31.39 / 0.855} & 32.87 / 0.868   \\ \hline
11                     & \multicolumn{1}{c|}{27.08 / 0.863} & \multicolumn{1}{c|}{25.35 / 0.842} & 28.38 / 0.877   & 14                     & \multicolumn{1}{c|}{32.18 / 0.857} & \multicolumn{1}{c|}{31.70 / 0.853} & 33.08 / 0.869   \\ \hline
12                     & \multicolumn{1}{c|}{28.01 / 0.875} & \multicolumn{1}{c|}{26.65 / 0.861} & 29.07 / 0.885   & 15                     & \multicolumn{1}{c|}{32.25 / 0.851} & \multicolumn{1}{c|}{31.75 / 0.845} & 33.00 / 0.869   \\ \hline
\end{tabular}
\end{table*} 
\begin{figure*}[t!]
    \centering
    \includegraphics[width=0.99\linewidth]{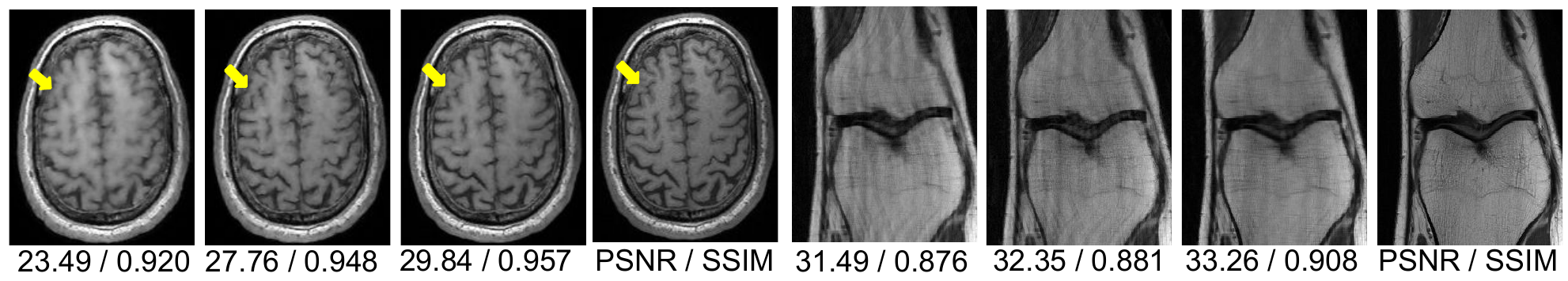}
    
    \caption{ Qualitative comparison of the HyperCoil-Recon with other adaptive MRI reconstruction architectures - MAC-ReconNet and AdaIN for a multimodal scenario that combines 7, 10, and 12 coils of T1 brain and PD knee datasets. From left: AdaIN, MAC-ReconNet, HyperCoil-Recon, and the GT images for the brain and knee anatomies.
    }
    \label{fig:mulanatomy}
\end{figure*}
\subsubsection{Expressivity in Multi-modal Scenario}
We combine multimodal contexts in a single model with multiple anatomies, PD knee and T1 brain, and multiple coil configurations (7, 10, and 12 coils), with 5x under-sampling.
 We compare our model with other adaptive MRI reconstruction  \cite{dc_cnn} architectures - MAC-ReconNet \cite{mac} and AdaIN \cite{univusmri}. Table \ref{tab:mulanatomy} and Figure \ref{fig:mulanatomy} show the results for the brain and knee for multiple coil configurations. We note that HyperCoil-Recon performs better than MAC-ReconNet and AdaIN for almost all the coil configurations both in PSNR and SSIM. MAC-ReconNet lacks task-invariant low-level image features as the dynamic weight prediction hypernetworks infer all the weights of the primary reconstruction network. On the other hand, AdaIN normalizes the features to compensate for the distribution shift using scale and shift operations. The hypernetworks in our model exhibit dense multiplicative interactions \cite{MultiplicativeIA} via convolution operations between predicted task-specific weights and the CNN features. These interactions provide comprehensive task-specific embeddings over the scale and bias operations of AdaIN and help to achieve mode-specific inductive bias \cite{kmmaml, multimodameta} when integrating diverse streams of contextual information in a single model.   

\subsubsection{Ablative Study}
\begin{figure}[t!]
    \centering
    \includegraphics[width=\linewidth]{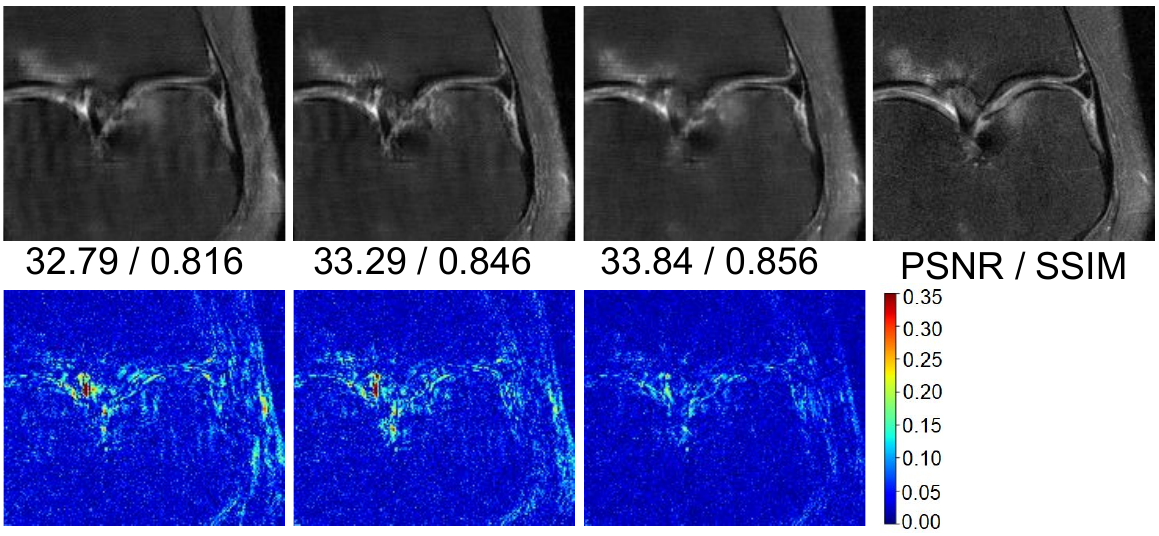}
    
    \caption{ An ablative study of cases with no DWP, DWP on the bottleneck layer and all the decoder layers. The residual images with respect to GT images are shown.
    }
    \label{fig:ablative}
\end{figure}

We perform an ablative study to understand the role of the hypernetworks on the top levels of the decoder and the bottleneck layer. 
We consider three cases across varying accelerations - 1) No dynamic weight prediction (DWP), 2) DWP only in the bottleneck layer, and 3) DWP in all layers \ie bottleneck and top layers of the decoder. Table \ref{tab:ablative}) shows the performance improvement with an increase in the number of hypernetworks as the contextual knowledge is improved. The qualitative results are shown in Figure \ref{fig:ablative}.

\begin{table}[t!]
\centering
\scriptsize
\caption{Ablative study for an increasing number of DWP hypernetworks in the bottleneck and top layers of the decoder across acceleration factors for coronal PDFS knee}
\label{tab:ablative}
\begin{tabular}{|l|c|c|}
\hline
\multicolumn{1}{|c|}{\multirow{2}{*}{Model}} & 5x           & 8x           \\ \cline{2-3} 
\multicolumn{1}{|c|}{}                       & PSNR/ SSIM   & PSNR/ SSIM   \\ \hline
No DWP                                       & 32.39/ 0.823 & 30.56/ 0.798 \\ \hline
DWP (only in bottleneck layer)                      & 33.00/ 0.890 & 30.93/ 0.861 \\ \hline
DWP (both in top \& bottleneck layers)                            & 33.65/ 0.893 & 31.34/ 0.862 \\ \hline
\end{tabular}
\end{table}

\section{Summary and Conclusion}

We introduce a unified coil-configuration task-switching CNN in a multi-tasking perspective to infuse the knowledge of dynamic coil configurations in multi-coil MRI reconstruction. Unlike conventional DL models, which scale with the number of coil configurations,  the proposed network is simple and parameter-efficient by design. With each configuration posed as a task, our model uses hypernetworks to infer task-specific weights that are embedded in the primary reconstruction network, enabling on-the-fly adaptation to multiple coil configurations. Our experiments demonstrate the efficacy of our approach in terms of high expressivity by interpolating to several unseen configurations,  better performance over other reconstruction methods, context-specific and context-invariant training methods, and insights into relationships between configurations. Our future direction is to extend the work for self-supervised learning for more PI settings in a single model.

{\small
\bibliographystyle{ieee_fullname}
\bibliography{cvamd2}
}

\end{document}